\documentclass[12pt,preprint]{aastex}

\usepackage{amsmath,amssymb}
\usepackage{epsfig}
\usepackage{graphicx}
\usepackage{natbib}
\usepackage{subfigure}
\usepackage{mathrsfs}
\usepackage{float}
\usepackage{array}
\usepackage{rotating}
\usepackage{bm}

\shorttitle{A coherent method for the detection and estimation of continuous gravitational wave 
            signals using a pulsar timing array}

\shortauthors{Wang et al.}

\begin{document}
    
\title{A coherent method for the detection and estimation of continuous gravitational wave 
            signals using a pulsar timing array}


\author{Yan Wang\altaffilmark{1,2}, Soumya D. Mohanty\altaffilmark{1,3} and Fredrick A. Jenet\altaffilmark{1,2,3}}
\affil{Department of Physics and Astronomy, University of Texas at
    Brownsville, TX 78520}
\affil{Center for Advanced Radio Astronomy, Brownsville, TX 78520}
\affil{Center for Gravitational Waves Astronomy, Brownsville, TX 78520}




\begin{abstract}
The use of a high precision pulsar timing array is a promising approach to detecting 
gravitational waves in the very low frequency regime ($10^{-6} -10^{-9}$ Hz) 
that is complementary to the ground-based efforts (e.g., LIGO, Virgo) at high frequencies 
($\sim 10 -10^3$ Hz) and space-based ones (e.g., LISA) at low frequencies 
($10^{-4} -10^{-1}$ Hz).  One of the target sources for pulsar timing arrays are 
individual supermassive black hole binaries that are expected to form in galactic 
mergers. In this paper, a likelihood based method for detection and estimation is 
presented for a monochromatic continuous gravitational wave signal emitted by such 
a source. The so-called pulsar terms in the signal that arise due to the breakdown 
of the long-wavelength approximation are explicitly taken into account in this method.
In addition, the method accounts for equality and inequality constraints 
involved in the semi-analytical maximization of the likelihood over a subset of the 
parameters. The remaining parameters are maximized over numerically using Particle 
Swarm Optimization. Thus, the method presented here solves the monochromatic 
continuous wave detection and estimation problem  without invoking some of the 
approximations that have been used in earlier studies.

\end{abstract}





\section{Introduction} \label{sec:intro}

Several worldwide projects are in progress to open the gravitational 
wave window for observational astronomy \citep{lrr-2009-2}. Taken together, 
these projects will span a wide range of astrophysically promising 
source frequencies, providing complementary views of the GW Universe. 
In the 10~Hz to 1~kHz range, the LIGO \citep{2009RPPh...72g6901A} and 
Virgo \citep{2011CQGra..28k4002A}  projects have already performed several 
joint observational runs \citep{2012PhRvD..85l2007A, 2012PhRvD..85h2002A}. 
Work is now in progress to commission second generation ground based 
detectors (Advanced LIGO \citep{2011arXiv1103.2728W}, 
Advanced Virgo \citep{2013ASPC..467..151D}, and KAGRA \citep{2012CQGra..29l4007S}). 
There have been long-standing plans for the 0.1 mHz to 0.1~Hz regime using 
space-based detectors. 
The LISA Pathfinder is scheduled for a launch in 2015. 
It will demonstrate and test the technologies to be used in the 
eLISA mission \citep{2013arXiv1305.5720C}.  
In the very low frequency 
regime of $10^{-9}$ to $10^{-6}$ Hz, the use of Pulsar Timing Arrays (PTA) 
is currently being explored intensively. 

In the past several decades, pulsar timing has produced many significant 
discoveries in astronomy which include the first evidence 
of the existence of gravitational waves \citep{1982ApJ...253..908T, 1989ApJ...345..434T}.
In pulsar timing \citep{2004hpa..book.....L}, time of arrivals (TOAs) of radio pulses from a 
rotating pulsar are fitted by a linearized timing model \citep{2006MNRAS.372.1549E}. 
The parameters of the model can be classified into astronomical 
(e.g. sky location, proper motion, period, period derivatives), interstellar medium (dispersion measure), 
binary system and instrumental parameters. The differences between the actual and the best-fit 
arrival times are called timing residuals. By subtracting out all known effects, the timing 
residuals from a given pulsar should just reduce to pure noise. Any deviation from this 
expectation may be attributed to the effect of a gravitational wave passing between the 
pulsar and Earth. 
 
The sensitivity of GW detection using pulsar timing is characterized by the root mean square 
(RMS) of the timing residuals. Currently a handful of millisecond pulsars (MSPs) have archived 
RMS at 100 ns level over many years \citep{2009MNRAS.400..951V, 2013ApJ...762...94D}. 
It has been shown that an array of pulsars timed to this level of precision 
can be operated as a galactic scale instrument to detect very low frequency 
GWs \citep{1983ApJ...265L..39H, 1990ApJ...361..300F}. A detection using such 
a PTA can be made by observing 20--40 pulsars over 5--10 years, assuming a monthly 
observation cadence and 100 ns timing precision for each \citep{2005ApJ...625L.123J}. 

At present, there are three major PTAs in operation: the North American 
Nanohertz Observatory for Gravitational Waves (NANOGrav, \citet{2013ApJ...762...94D}), 
the Parkes Pulsar Timing Array (PPTA, \citet{2013PASA...30...17M}), 
and the European Pulsar Timing Array (EPTA, \citet{2010CQGra..27h4014F}). 
As independent consortia, the three PTAs compose the International Pulsar Timing 
Array (IPTA, \citet{2013arXiv1309.7392M}) with approximately 50 pulsars regularly monitored. 
Data shared between PTAs can form a longer observation duration and finer 
observation cadence for a specific pulsar (e.g. PSR J1713+0747, \citet{2013inPrepZhu}). 
In addition, combining geographically widely distributed telescopes 
of all the PTAs increases sky coverage and reduces certain systematics. 

One of the principle signals anticipated in PTA based GW detection is the stochastic 
background formed by the incoherent superposition of GW signals from a large 
population of unresolved sources distributed throughout the Universe. 
Besides this background, it may be possible to detect and characterize 
individual sources using a PTA \citep{2010PhRvD..81j4008S}. Using the data 
from the Millennium Simulation of structure formation \citep{2005Natur.435..629S}, 
and considering a broad range of population models for Super Massive Black Hole Binaries (SMBHBs), 
\citet{2009MNRAS.394.2255S} found that sufficiently strong GW signals generated by close and/or 
massive SMBHBs could stand above the stochastic background and be resolved individually. 
Pulsar timing has already been used to rule out the radio galaxy 3C 66B \citep{2003Sci...300.1263S}
as SMBHB candidate due to the absence of any detectable effect on pulsar timing 
residuals by the GW signal that should have been produced by a binary source \citep{2004ApJ...606..799J}. 
Besides SMBHBs, intermediate-mass black hole binaries that may reside in globular clusters 
could also be potential GW sources for PTA \citep{2005ApJ...627L.125J}. 

Along with observational advances in pulsar timing precision, 
there are data analysis issues that must be satisfactorily 
resolved in order to increase the sensitivity of PTA-based GW 
detection. PTA data presents several features, such as irregular 
sampling and non-stationary noise, that data analysis algorithms 
must contend with. Since different algorithms may react differently 
to these features, it is important that a variety of algorithms 
be developed and compared. Several different data analysis methods 
have been proposed for the case of continuous GW signals 
\citep{2010MNRAS.407..669Y, 2010arXiv1008.1782C, 2012PhRvD..85d4034B, 
2012ApJ...756..175E, 2013CQGra..30v4004E}. 

It is natural that methods for continuous source detection using PTA 
share much in common with those already in use for ground-based detectors. 
Following the Generalized Likelihood Ratio Test (GLRT) approach \citep{1998.book.....KayII}, 
\citet{1998PhRvD..58f3001J} developed the $\mathcal{F}$-statistic method 
for a single detector with a known noise model. This method was later generalized by 
\citet{2005PhRvD..72f3006C} to the case of a network 
of detectors with time-varying noise.  In the PTA regime, the long wavelength 
approximation that is used to obtain the $\mathcal{F}$-statistic is no longer valid. 
This results in the phase of the gravitational wave at a pulsar being substantially different 
from the one at Earth. Hence, the timing residual signal due to a GW source acquires 
additional parameters related to the distance from the pulsar to Earth and the direction 
of the source relative to the line of sight to the pulsar. There are as many of these 
additional so-called pulsar phase parameters brought into the maximization 
of the likelihood as there are pulsars in a PTA. 

In the case of the $\mathcal{F}$-statistic, the likelihood is analytically 
maximized over a subset of signal parameters. These parameters are called 
extrinsic, while the ones left over are called intrinsic. The division of 
the signal parameters into these subsets is quite natural and unique. However, 
once the pulsar phase parameters appear in the picture, this division is 
no longer unique. One option is to choose the same division as is done for 
the $\mathcal{F}$-statistic. In this case, the pulsar phases have to be 
maximized numerically. The other option is to treat the pulsar phases as 
extrinsic, but this pushes all other parameters into the class of intrinsic 
parameters. The latter option may well be the direction that needs to be followed, 
especially when it comes to a PTA with a large number of pulsars or the 
simultaneous detection and estimation of multiple sources, 
but it has not been well explored yet. The former option is better understood, 
but it increases the computational cost of maximizing the likelihood, due to 
the large number of additional parameters, by orders of magnitude. 

Previous analyses of PTA sensitivity to individual SMBHB signals have 
ignored the pulsar phase parameters \citep{2012PhRvD..85d4034B}, or handled 
them suboptimally \citep{2012ApJ...756..175E}. However, \citet{2010arXiv1008.1782C} 
showed that excluding these parameters can drastically reduce sensitivity as well 
as increase parameter estimation errors. They derive this conclusion from a 
Bayesian approach where a single data realization was used to derive statistical 
measures of error from the posterior degree of belief \citep{2010blda.book.....G}. 
It was also found that an estimation of 
the pulsar phase parameters could lead to reasonably accurate (compared with 
parallax measurements derived) estimates of the distances to some of the pulsars in a PTA.

In this paper, we investigate the efficacy of GLRT detection and estimation 
following the option of treating the pulsar 
phase parameters as intrinsic. It is shown that the analytic maximization over 
the extrinsic parameters is actually a constrained optimization problem involving an 
equality and an inequality constraint, both of which are non-linear. Our 
proposed method takes these constraints into account explicitly. To overcome 
the computational barrier posed by the above choice of intrinsic parameters, 
we use Particle Swarm Optimization (PSO). This optimization method was 
introduced by \citet{eberhart1995new} and has become 
quite popular across a wide range of fields. PSO was introduced in GW data 
analysis by \citet{2010PhRvD..81f3002W}. Its use in PTA has been explored by 
\citet{2012arXiv1210.3489T}, although it was only employed for a two-dimensional 
search space. The method presented here uses PSO over a 12-dimensional search space.

The performance of our method is investigated using three test cases spanning 
a wide range of signal strengths from the very strong to the barely detectable. 
(Quantified using the network signal to noise ratio defined in Sec. \ref{sec:results}, 
the values used are 100, 10 and 5.)  We use a large number of independent data 
realizations and derive conventional Frequentist error estimates for the signal 
parameters. The strong signal case shows that the estimation of the signal 
waveform in the timing residual of each pulsar has a weak dependence on the 
estimation accuracy of pulsar phase parameters. This suggests that the pulsar 
phases are actually nuisance parameters and that the likelihood is highly 
degenerate when considered as a function over these parameters. 
As the signal strength is reduced to more realistic levels, the estimated pulsar 
phase parameters acquire a nearly uniform distribution across their allowed range. 
Therefore, this suggests strongly that in future algorithms, they should be treated 
as extrinsic parameters (or as parameters that are marginalized over). 
Within the domain of the approximations made in our study, we find that it should 
be possible to make a confident detection of an SMBHB signal at astrophysically 
realistic signal strengths. As an example, for an SMBHB of $10^9~M_{\odot}$ chirp mass 
with an orbital period of $0.785$ years located at $100$ Mpc, for which the observed 
signal has a network matched filtering signal to ratio of 10, the detection 
probability is $\approx 1$ at a false alarm probability $\sim 10^{-3}$. 
The one-sigma contour of the estimated sky location encloses an area of 
800 $\text{deg}^2$, but the orbital frequency can still be estimated with 
a standard deviation $<0.1~\text{rad~yr}^{-1}$.

The rest of the paper is organized as follows. In Sec.~\ref{sec:san} we introduce the data model 
used in this paper. Sec.~\ref{sec:Fstats} describes the GLRT for this data model which involves 
constrained maximization over extrinsic parameters. Sec.~\ref{sec:pso} presents a brief review of 
PSO and discusses its implementation in this application. Sec.~\ref{sec:results} demonstrates 
the method and quantifies its performance using test cases. The paper is concluded in 
Sec.~\ref{sec:sum}.

\section{Data model} \label{sec:san}

The data from a PTA consists of a set of sequences of timing residuals, 
one for each pulsar, $r^I = (r^I_1,r^I_2,\ldots,r^I_{N_I})$, $I=1,2,\ldots,N_p$, 
where $N_p$ is the number of pulsars, $N_I$ is the number of observation for the 
$I$-th pulsar, and $r^I_i$ is the timing residual 
observed at time $t^I_i\in [0,T]$, $t^I_{i+1}>t^I_i$, by an observer at 
Earth. 
\begin{eqnarray}
r^I_k &=& s^I_k+n^I_k ~~\text{when a GW signal is present}  \,,   \\
      &=& n^I_k ~~\text{when a GW signal is absent.}
\end{eqnarray}\label{eq:datamodel}
where $n^I=(n^I_1,n^I_2,\ldots,n^I_{n_I})$ and $s^I = (s^I_1,s^I_2,\ldots,s^I_{n_I})$ 
denote the noise realization and the GW signal in the $I$-th timing residual sequence.
In this section, we first introduce the GW signal family used in this paper, followed by
a description of the noise model.

\subsection{Timing residuals induced by GW} \label{sec:Fstats1}

%
\begin{figure}[h]
\centerline{\includegraphics[width=0.4\textwidth]{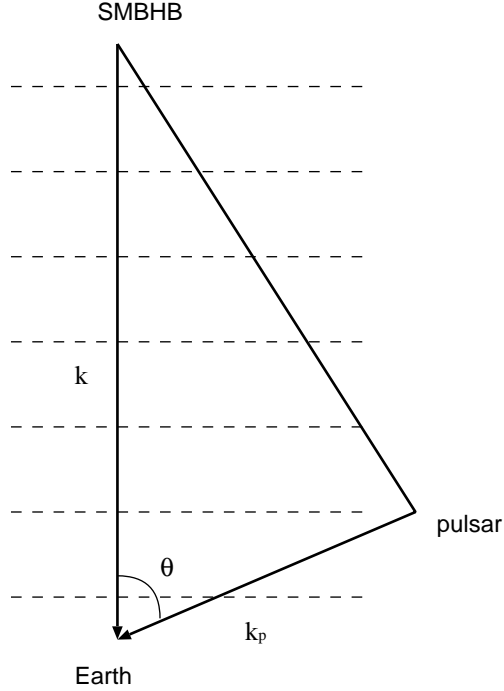}}
\caption{Supermassive black hole binary, pulsar and Earth configuration. $\theta$ is 
the open angle between the supermassive black hole binary and the pulsar subtended at Earth. 
$\mathbf{k_p}$ is the wave vector of a radio pulse, $\mathbf{k}$ is the wave vector 
of GWs. Dash lines represent the wave front of the GWs.}
\label{fig:pulsarblackhole}
\end{figure}
%

Consider the scenario shown in Fig.~\ref{fig:pulsarblackhole} where 
plane GWs from some distant continuous source, such as a SMBHB, are crossing 
the line of sight from Earth to a Galactic pulsar. The stretching and 
squeezing of the path length of radio pulses by the GW cause fluctuations 
in the arrival time of the pulses at Earth. Even though individual radio 
pulses from a pulsar are emitted irregularly in time, integrated pulses 
have a predictable TOA \citep{2004hpa..book.....L}. 
Any deviation from the predicted TOAs is called the timing residual, part 
of which may be attributed to the effect of a GW. The effect of the GW 
signal on the timing residual 
is achromatic, i.e. the magnitude of the fluctuation is independent of the carrier 
frequency of the radio emission. The fluctuation of a specific pulse can 
be calculated by integrating the fractional change in frequency 
of the pulse rate (time redshift) induced by GW along the trajectory from 
the pulsar to the observer. The GW tensor is 
\begin{equation}
\mathbf{h}=(A_+ \mathbf{e}_+ + A_{\times} \mathbf{e}_{\times}) e^{i(\omega_{gw}t-\mathbf{k}\cdot\mathbf{x})} \,,  
\end{equation}\label{gw}
where $\omega_{gw}$ is the angular frequency, $\mathbf{k}$ is the 
wave vector of GW. For a non-evolving binary with circular orbit, 
which is considered in this paper, $\omega_{gw}=2\omega$ ($\omega$ is the angular frequency of the binary). 
$\mathbf{e}_+ = \boldsymbol{\hat{\alpha}}\otimes\boldsymbol{\hat{\alpha}} 
-\boldsymbol{\hat{\delta}}\otimes\boldsymbol{\hat{\delta}}$ and 
$\mathbf{e}_\times = \boldsymbol{\hat{\alpha}}\otimes\boldsymbol{\hat{\delta}}
+\boldsymbol{\hat{\delta}}\otimes\boldsymbol{\hat{\alpha}}$ are the basis 
tensors of $+$ and $\times$ polarizations, and $\boldsymbol{\hat{\alpha}}$ 
and $\boldsymbol{\hat{\delta}}$ are the basis vectors of the right ascension and 
declination of the equatorial coordinates. Expressions for $A_{+}$ and $A_{\times}$ can be found in 
\citet{2007.book.....M}.  Without loss of generalization, it is assumed 
in the following that the pulsar and observer are both fixed 
in an inertial reference frame (their motions relative to this 
reference frame can be treated separately as an additive effect). 
For $I$-th pulsar, the timing residual induced by a monochromatic continuous 
GW signal for an observer at Earth is 
\begin{eqnarray}
s^I_k &=& \int_{0}^{t^I_k} z(t')\, \mathrm{d}t' = s_e^I(t^I_k)-s_p^I(t^I_k)  \,, \nonumber \\
     &=& \Re\bigg\lbrace \frac{1}{2i}\cdot\frac{\mathbf{k^\text{I}_p}\cdot \mathbf{A} \cdot 
     \mathbf{k^\text{I}_p}}{1-\cos\theta^I}\cdot e^{i (2\varphi_0+\omega_{gw}t^I_k)} 
     \Big(1-e^{-i\omega_{gw}d_p^I(1-\cos\theta^I)}\Big) \bigg\rbrace \,.  \label{eq:resint}
\end{eqnarray}
where $z(t')$ is the redshift of the pulse at time $t'$, $s_e^I$ and $s_p^I$ are 
the so called Earth term and pulsar term respectively, $\Re$ is the real part 
of the expression in the parenthesis, $\mathbf{k^\text{I}_p}$ is the wave vector of 
pulse, $\varphi_0$ is the initial phase of the binary orbit at the starting time 
of the observations, $\theta^I$ is the open angle between the SMBHB and the 
pulsar subtended at the Earth, and $d_p^I$ is the distance between the pulsar and 
Earth.  

The signal can be conveniently rewritten as,
\begin{eqnarray}
s^{I}_k = \sum_{\mu=1}^{4} a_{\mu}A_{\mu}^{I}(t^I_k)  \,. \label{eq:residual2}
\end{eqnarray}
The coefficient $a_\mu$ are given by 
\begin{subequations}\label{eq:lowa}
\begin{align}
a_1=-\zeta(1+\cos^2\iota)\cos 2\psi  \label{first}   \,, \\
a_2=-2\zeta\cos\iota\sin 2\psi  \label{second}   \,, \\
a_3=\zeta(1+\cos^2\iota)\sin 2\psi  \label{fifth}   \,, \\
a_4=-2\zeta\cos\iota\cos 2\psi  \label{sixth}   \,, 
\end{align}
\end{subequations}
where $\iota$ is the inclination angle between the binary orbital plane 
and the plane of the sky, $\psi$ is the GW polarization angle, and 
\begin{equation}\label{eq:zeta}
\zeta=\frac{G\mu a^2\omega^2}{c^4 D}=\frac{G^{5/3}}{c^4 D}\mathcal{M}_c^{5/3}\omega^{-1/3} \,,
\end{equation}
where $\mu$ is the reduced mass, $\mathcal{M}_c=\mu^{3/5}M^{2/5}$ is the chirp mass 
of the binary source, $M$ is its total mass, $a$ is its semi-major axis, and $D$ is the 
distance from Earth to the source. These coefficients only contain the source parameters 
$\zeta, \iota, \psi$ which are shared among all pulsars. 

The basis functions $A_{\mu}^{I}(t^I_k)$ are 
\begin{subequations}\label{eq:capitala}
\begin{align}
A^I_1(t^I_k)=-2F^I_{+}\sin(\varphi_0-\varphi_I)\sin(\varphi_0+\varphi_I+\Phi(t^I_k))   \label{first}  \,, \\
A^I_2(t^I_k)=2F^I_{+}\sin(\varphi_0-\varphi_I)\cos(\varphi_0+\varphi_I+\Phi(t^I_k))  \label{second}  \,, \\
A^I_3(t^I_k)=-2F^I_{\times}\sin(\varphi_0-\varphi_I)\sin(\varphi_0+\varphi_I+\Phi(t^I_k))  \label{third}  \,, \\
A^I_4(t^I_k)=2F^I_{\times}\sin(\varphi_0-\varphi_I)\cos(\varphi_0+\varphi_I+\Phi(t^I_k))  \label{fourth}  \,,
\end{align}
\end{subequations}
where the parameters are: the right ascension $\alpha$ and 
declination $\delta$ of the source, $\omega_{gw}$, $\varphi_0$, and the pulsar phase $\varphi_I$ 
(considering that the right ascension $\alpha^I_p$ and declination $\delta^I_p$ of the pulsar 
are known with high precision.) 
Note that $\varphi_I=\varphi_0-\omega d^I_p(1-\cos\theta^I)$ defined for pulsar $I$ will be 
regarded as an independent parameter in this analysis. $\Phi(t)$ is the phase evolution of 
the GWs. For circular orbit, $\Phi(t^I_k)=\omega_{gw}\,t^I_k$. Here $F^I_+=P^I_+/(1-\cos\theta^I)$ 
and $F^I_\times=P^I_\times/(1-\cos\theta^I)$ are the \textit{antenna pattern functions} 
with $P^I_+$ and $P^I_\times$ are defined as ($\tilde{\alpha}^I=\alpha-\alpha^I_p$)
\begin{equation}\label{eq:antennapattern1}
P^I_+=-\cos^2\delta^I_p(1-2\cos^2\tilde{\alpha}^I+\cos^2\tilde{\alpha}^I\cos^2\delta)
   +\sin^2\delta^I_p\cos^2\delta -\sin 2\delta^I_p\cos\tilde{\alpha}^I\sin\delta\cos\delta  \,,
\end{equation}
and
\begin{equation}\label{eq:antennapattern2}
P^I_\times=2\cos\delta^I_p\sin\tilde{\alpha}^I(-\sin\delta^I_p\cos\delta+\cos\delta^I_p\cos\tilde{\alpha}^I\sin\delta)  \,.
\end{equation}

For a supermassive black hole binary, the overall amplitude is
\begin{equation}\label{eq:amp}
\zeta\approx 5\times 10^{-7} \left(\frac{\mathcal{M}_c}{10^9~M_\odot}\right)^{5/3} 
\left(\frac{D}{10~\text{Mpc}}\right)^{-1} \left(\frac{P}{5~\text{yr}}\right)^{1/3} \text{sec}  \,.
\end{equation}
where $P=2\pi/\omega$ is the orbital period of the binary.

\subsection{The noise model} \label{sec:Fstats2}

It is commonly assumed in studies of PTA data analysis methods that the 
noise process in timing residuals is Gaussian and stationary. The probability 
density function of the noise time series can be characterized by 5 parameters 
for each pulsar \citep{2014arXiv1404.1267A}. 

One way to account for the noise parameters, as is done for the $\mathcal{F}$-statistic, 
is to estimate them independently and use the estimates as fixed values in the detection 
and estimation of signals. This approach is a good approximation when the signal 
is weak and has no significant effect on the noise parameter estimates.  Another approach 
is to include the noise parameters along with those of the signal in an overall estimation 
method. This approach leads to significantly higher computational costs and its use 
in the weak signal case must be examined with care as it may not lead to substantial 
improvement in the final results. 

In this paper, following previous works \citep{2012PhRvD..85d4034B, 2012ApJ...756..175E, 2014arXiv1404.1267A}, 
we assume that the noise processes in the $N_p$ pulsars are mutually independent Gaussian 
and stationary, and use the $\mathcal{F}$-statistic approach of using independently 
estimated noise parameters. 

Define the \textit{noise weighted inner product} for two time series $x$ and $y$ 
\begin{equation}\label{eq:noiseprod}
\langle x|y \rangle_I =x \Sigma_I^{-1} y^T \,,
\end{equation}
where $\Sigma_I=\text{E}(n^In^I)$ is the auto-covariance matrix of the noise 
process in the pulsar $I$. Then the joint probability density function 
of the data in the absence of any GW signal is given by 
\begin{equation}\label{eq:gauss2}
p(\mathbf{n})=\prod_{I=1}^{N_p} p(n^I)= \prod_{I=1}^{N_p}\frac{1}{(2\pi)^{N_I/2}|\Sigma_I|^{1/2}}
\exp\left[-\frac{1}{2}\langle n^I|n^I\rangle_I \right]  \,. 
\end{equation}
Here $\bf n$ denotes the set $\lbrace n^1, n^2,...,n^{N_p}\rbrace$. Each $n^I$ 
is a row vector representing the noise time series for the $I$-th pulsar, and its length is 
determined by the number of observations $N_I$ conducted for this pulsar. We have assumed that $n^I$ 
are mutually independent between pulsars. $|\Sigma_I|$ is the determinant of the auto-covariance matrix.

\section{Maximum likelihood ratio statistic} \label{sec:Fstats}

In hypotheses testing, we formulate two mutually exclusive hypotheses 
for a pulsar timing array data set: 

$\bullet$ $\mathcal{H}_0$: $r^I(t)=n^I(t)$, $I=1,\ldots, N_p$, there is no signal in the data.

$\bullet$ $\mathcal{H}_\lambda$: $r^I(t)=n^I(t)+s^I(t;\lambda)$, $I=1,\ldots, N_p$, there is a signal 
characterized by parameter $\lambda$ in the data. 

\noindent Here $\mathcal{H}_\lambda$ is a \textit{composite} hypothesis, and 
$\lambda=\lbrace\zeta, \iota, \psi, \alpha, \delta, \omega_{gw}, \varphi_0, \varphi_I \rbrace$ 
($I=1,2,\ldots, N_p$).  
One needs to choose between these two hypotheses based on the observed data. 
Geometrically, it is equivalent to divide the observation space $R^N$ ($N$-dimensional 
real space, $N=\sum_{I=1}^{N_p}N_I$) into 
two disjoint regions $R_0$ and $R_\lambda$. If the data $\textbf{r}=(r^1,\ldots, r^{N_p})\in R_\lambda$, $\mathcal{H}_\lambda$ is 
chosen; while if the data $\textbf{r}\in R_0$, $\mathcal{H}_0$ is chosen. The boundary of the two regions is called 
the \textit{decision surface} $D$. Different detection strategies are distinguished 
by different choices of $D$. In general, there does not exist a $D$ that is optimal 
for all values of $\lambda$. Only in the special case where $\lambda$ is completely 
known (or some trivial extensions thereof) does one get an optimal $D$. Following 
the \textit{Neyman-Pearson criterion}, in which $D$ minimizes false dismissal probability 
for a given false alarm probability, the optimal $D$ turns out to be an iso-surface of 
the likelihood ratio, 
\begin{equation}\label{eq:lr}
\text{LR}(\mathbf{r})=\frac{p(\mathbf{r}|H_\lambda)}{p(\mathbf{r}|H_0)} \,.
\end{equation}
In the case of an unknown $\lambda$, a natural modification 
is to use iso-surfaces of the GLRT functional of data $\mathbf{r}$ obtained by maximizing the 
likelihood ratio over $\lambda$, 
\begin{equation}\label{eq:glrt}
\text{GLRT}(\mathbf{r})=\max_{\lambda}\frac{p(\mathbf{r}|H_\lambda)}{p(\mathbf{r}|H_0)} \,.
\end{equation}
Note that the same decision surface can be obtained 
by replacing the likelihood ratio by any monotonic function, such as the logarithm, 
of itself. In the following we use the maximum of $\Lambda(\mathbf{r})=\ln(\text{LR}(\mathbf{r}))$ 
over $\lambda$ as the detection statistic.

\subsection{The likelihood ratio} \label{ModF}

The logarithm of the likelihood ratio of the hypotheses in Eq.~\ref{eq:lr} is 
\begin{equation}\label{eq:loglambda}
\Lambda(\mathbf{r})=\ln \frac{p(\mathbf{r}|\mathcal{H}_\lambda)}{p(\mathbf{r}|\mathcal{H}_0)}
=\sum_{I=1}^{N_p}\langle r^I|s^I(\lambda)\rangle_I  -\sum_{I=1}^{N_p}\frac{1}{2}\langle(s^I(\lambda)|s^I(\lambda)\rangle_I \,. 
\end{equation}
The detection statistic is the maximum of $\Lambda(\mathbf{r})$
over the parameter space $\lambda$.  Although the number of parameters 
involved in the maximization of $\Lambda(\mathbf{r})$ are large, a subset of 
them can be maximized efficiently using a semi-analytical approach. 
However, the choice of this subset is not unique. Following the approach 
used in the $\mathcal{F}$-statistic, we could pick $\zeta, \iota, \psi$ 
as the subset, which would require a numerical maximization over the 
pulsar phase $\varphi_I$ and other parameters in the complementary subset. 
Alternatively, one could choose the $\varphi_I$ as the subset 
leading to numerical maximization over the complementary subset. 
The merits and demerits of these two alternative approaches need much 
further study. Here, we simply follow the $\mathcal{F}$-statistic approach 
and divide the parameters into a subset $\lambda_e=\lbrace\zeta, \iota, \psi\rbrace$, 
to be maximized semi-analytically, and a subset 
$\lambda_i=\lbrace\alpha, \delta, \omega_{gw}, \varphi_0, \varphi_I \rbrace$ 
to be maximized numerically.

For a network of $N_p$ pulsars, and using Eq.~\ref{eq:residual2} 
we have from Eq.~\ref{eq:loglambda} 
\begin{eqnarray}\label{eq:lognetwork}
\Lambda(\mathbf{r}) 
&=& \sum_{\mu=1}^{4}a_{\mu}\sum_{I=1}^{N_{p}} \langle r^I|A_\mu^I \rangle_I-\frac{1}{2}
\sum_{\mu=1}^{4}\sum_{\nu=1}^{4}a_\mu a_\nu\sum_{I=1}^{N_{p}}\langle A_\mu^I|A_\nu^I\rangle_I \\ 
&=& \mathbf{a}\cdot\mathbf{N}-\frac{1}{2}\mathbf{a}\cdot\mathbf{M}\cdot\mathbf{a}
\end{eqnarray}
where $N_{\mu}=\sum_{I=1}^{N_p} \langle r^I|A_\mu^I \rangle_I$ is an $4\times 1$ vector which
contains the data and the intrinsic parameters $\lbrace\alpha, \delta, \omega_g, \varphi_0, \varphi_I \rbrace$, 
$M_{\mu\nu}=\sum_{i=1}^{N_p}\langle A_\mu^I|A_\nu^I\rangle_I$ is an $4\times4$ matrix that 
contains the intrinsic parameters only. The unconstrained maximum \citep{2012PhRvD..85d4034B,2012ApJ...756..175E}
of $\Lambda(\mathbf{r})$ is easily obtained from, 
\begin{equation}\label{eq:maxloglike}
\frac{\partial \Lambda}{\partial \mathbf{a}}= \mathbf{N} - \mathbf{M}\cdot\mathbf{a} = \mathbf{0} \,,
\end{equation}
for which the solution is $\mathbf{a}=\mathbf{M}^{-1}\cdot\mathbf{N}$. Then the 
maximum of $\Lambda(\mathbf{r})$ over the extrinsic parameters is 
\begin{equation}\label{eq:maxloglike2}
\max_{\lambda_e} \lbrace \Lambda\rbrace = \frac{1}{2}\mathbf{N}\cdot\mathbf{M}^{-1}\cdot\mathbf{N}\,. 
\end{equation}
Having solved for $\mathbf{a}$, the extrinsic parameters can be explicitly 
derived from Eq.~\ref{eq:lowa},
\begin{subequations}\label{eq:extrin}
\begin{align}
\zeta=\frac{1}{2}\Big(\big(a_1^2+a_3^2\big)^{1/2}+\big(a_1^2-a_2^2+a_3^2-a_4^2\big)^{1/2}\Big)  \label{first}  \,, \\
\psi=\frac{1}{2}\arctan\big(-a_3/a_1\big) \label{second}  \,,  \\
\iota=\arccos\big(-a_2/(2\zeta\sin\psi)\big) \label{third}  \,,
\end{align}
\end{subequations}
where $\zeta>0$, $\psi\in[0,\pi]$, and $\iota\in[0,\pi]$.

\subsection{Karush-Kuhn-Tucker conditions}\label{sec:kkt}

The unconstrained maximum presented above is not the correct solution 
because there are indeed additional constraints on $\mathbf{a}$. 
From Eq.~\ref{eq:lowa} we notice the fact that $\tan2\psi=-a_3/a_1=a_2/a_4$, 
therefore we have one nonlinear equality constraint (NEC), 
\begin{equation}\label{eq:constraint2}
a_1a_2+a_3a_4=0   \,. 
\end{equation}
It turns out that this condition also guarantees that the absolute value 
of the argument of the $\arccos$ function in Eq.~\ref{eq:extrin} 
is not greater than one. Furthermore, to have a meaningful (square root 
of a nonnegative number) solution of $\zeta$ in Eq.~\ref{eq:extrin} 
one additional nonlinear inequality constraint (NIEC) 
\begin{equation}\label{eq:nonconstraint}
a_1^2-a_2^2+a_3^2-a_4^2\geq 0   \,,
\end{equation}
is required. (Notice the $\arctan$ function is always meaningful.)

The \textit{Karush-Kuhn-Tucker} (KKT) conditions provide a formal framework 
for solving constrained optimization problems that include equality and inequality 
constraints. Essentially, these conditions state that the solution is guaranteed 
to lie in the region where an inequality constraint 
is satisfied or on the boundary of this region. The boundary is obtained by turning the 
inequality constraint into an equality constraint. 

Following the KKT prescription, the strategy for finding the solution of our problem 
is composed of two steps, as described below. The first step finds a solution that satisfies 
only the NEC. The solution to the second steps satisfy both NEC and NIEC. Only if the solution from 
the first step does not satisfy NIEC, is the second step taken.

Step 1: taking NEC into account.  We can include the nonlinear equality constraint 
in the Lagrangian method as follows 
\begin{equation}\label{eq:nonlllf}
\Lambda=\mathbf{a}^{T}\mathbf{N}-\frac{1}{2}\mathbf{a}^{T}\mathbf{M a}
+\frac{\mu}{2} \mathbf{a}^{T}\mathbf{D a}  \,,
\end{equation}
where $\mu$ is the Lagrangian multiplier for the nonlinear constraint, and 
\begin{equation}\label{Dmatrix}
\mathbf{D}=\left( \begin{array}{cccc}
0 & 1 & 0 & 0 \\
1 & 0 & 0 & 0 \\
0 & 0 & 0 & 1  \\
0 & 0 & 1 & 0 \\ 
\end{array} \right)\,.
\end{equation}
Then, differentiating Eq.~\ref{eq:nonlllf} with respect to $\mathbf{a}$
\begin{equation}\label{eq:partiala}
\frac{\partial \Lambda}{\partial \mathbf{a}}= \mathbf{N} -
\mathbf{M}\cdot\mathbf{a}+\mu\mathbf{D\cdot a} = \mathbf{0} \,,
\end{equation}
and the multiplier $\mu$ 
\begin{equation}\label{eq:partialmu}
\frac{\partial \Lambda}{\partial \mu}= \frac{1}{2}\mathbf{a\cdot D\cdot a} = 0 \,,
\end{equation}
and solving Eq.~\ref{eq:partiala} and Eq.~\ref{eq:partialmu}, we get 
\begin{equation}\label{eq:nonla}
\mathbf{a}_\text{NEC}=(\mathbf{M}-\mu \mathbf{D})^{-1}\mathbf{N} \,,
\end{equation}
where $\mu$ can be solved by using one dimensional numerical root finding on 
\begin{equation}\label{eq:nonlsolvemu}
\mathbf{N}^{T}(\mathbf{M}-\mu \mathbf{D})^{-1}\mathbf{D}
(\mathbf{M}-\mu\mathbf{D})^{-1}\mathbf{N}=0 \,.
\end{equation}
Having obtained $\mu$, we insert it back into Eq.~\ref{eq:nonla} to get $\mathbf{a}$ 
and then $\Lambda$ from Eq.~\ref{eq:lognetwork}. We then check if this solution of 
$\mathbf{a}$ violates the nonlinear inequality condition in Eq.~\ref{eq:nonconstraint}. 
If not, it is already the correct solution; if yes, we then need to 
incorporate this quadratic inequality in the Lagrangian multiplier equation. 

Step 2: taking NEC and NIEC into account. Since the solution of 
$\mathbf{a}$ always appears at the boundary of the region satisfying 
the inequality, this constraint reduces to $a_1^2-a_2^2+a_3^2-a_4^2= 0$. 
As above we need to solve the following problem 
\begin{equation}\label{eq:2nonlllf}
\Lambda=\mathbf{a}^{T}\mathbf{N}-\frac{1}{2}\mathbf{a}^{T}\mathbf{M a}
+\frac{\mu}{2} \mathbf{a}^{T}\mathbf{D a} 
+ \nu \mathbf{a}^{T}\mathbf{B a} \,,
\end{equation}
where $\nu$ is the Lagrangian multiplier that takes care of the second quadratic condition, and 
\begin{equation}\label{Bmatrix}
\mathbf{B}=\left( \begin{array}{cccc}
1 & 0 & 0 & 0 \\
0 & -1 & 0 & 0 \\
0 & 0 & 1 & 0 \\
0 & 0 & 0 & -1 \\ 
\end{array} \right)\,.
\end{equation}
As before, taking the partial derivatives of $\Lambda$ with respect to $\mathbf{a}$, 
$\mu$, and $\nu$ will get 
\begin{equation}\label{eq:2nonla}
\mathbf{a}_\text{NEC+NIEC}=(\mathbf{M}-\mu\mathbf{D}-2\nu\mathbf{B})^{-1}\mathbf{N} \,.
\end{equation}
Here multiplier $\mu$ and $\nu$ can be solved simultaneously from the following two equations
\begin{equation}\label{eq:nonlsolvemunu1}
\mathbf{N}^{T}(\mathbf{M}-\mu \mathbf{D}-2\nu\mathbf{B})^{-1}\mathbf{D}(\mathbf{M}
-\mu\mathbf{D}-2\nu\mathbf{B})^{-1}\mathbf{N}=0 \,,
\end{equation}
\begin{equation}\label{eq:nonlsolvemunu2}
\mathbf{N}^{T}(\mathbf{M}-\mu\mathbf{D}-2\nu\mathbf{B})^{-1}\mathbf{B}
(\mathbf{M}-\mu\mathbf{D} -2\nu\mathbf{B})^{-1}\mathbf{N}=0 \,.
\end{equation}
To solve the above two equations simultaneously one needs a two-dimensional numerical 
root finding method. This step is, therefore, computationally more demanding than the one-dimensional 
method in step 1. Fortunately the chance of facing this case is very small in practice, 
and our strategy to handle it is to numerically solve the quadratically-constrained 
quadratic program (QCQP) with the active-set algorithm \citep{2006no..book.....N} or 
the interior-point algorithm \citep{1999siam.9.877B}. We choose the former in 
our demonstration in Sec.~\ref{sec:results} for its computational efficiency, although 
the latter could be more accurate if the gradients and the Hessians of $\Lambda$ and 
the constraints are provided. Either algorithm needs a starting point for the search, 
and the $\mathbf{a}_\text{NEC}$ in step 1 is proved to be a good choice.

\section{Particle Swarm Optimization (PSO)}\label{sec:pso}

In Sec.~\ref{sec:kkt}, we showed how to maximize the extrinsic parameters $\lambda_e$ 
for a given set of intrinsic parameters $\lambda_i$. Maximization over 
$\lambda_i$ requires a search over a $4+N_p$ dimensional space. Due to the presence of noise and degeneracies in the intrinsic signal parameter space, $\Lambda(\mathbf{r})$ is a highly multi-modal function having a forest of local optima. Deterministic local optimization methods cannot
be used to locate the global optimum of such a function.  A brute force search using a grid of points is computationally prohibitive because the density of the grid must be high to tackle the large number of local optima while, at the same time, the number of grid points grows exponentially with the dimensionality of the search space.

The only feasible approach
when it comes to multi-modal and high dimensional optimization problems
is to use algorithms that employ some type of a stochastic search scheme. 
A large class of these algorithms are modeled after biological systems~\citep{engelbrecht2005fundamentals} with
Particle Swarm Optimization (PSO)~\citep{eberhart1995new} being among the popular choices. 
The use of PSO in gravitational wave data analysis was introduced in \citet{2010PhRvD..81f3002W}  where it was applied
to compact binary inspiral searches for ground based detectors. In~\cite{2012AstRv...7b..29M,2012AstRv...7d...4M}, PSO was applied to
the problem of GW burst detection where global optimization over a
high dimensional search space is required.
PSO has previously been used in
pulsar timing analysis for the case of a two dimensional search space 
\citep{2012arXiv1210.3489T, 2013PhRvD..87j4021L}. Here, we use PSO
in the context of the ($4+N_p$) dimensional search space. A brief description of PSO follows.

Consider the global maximization of a scalar function $f(x)$,
called the {\em fitness } function, where
$x\in S \subset \mathbb{R}^n$.  $S$ is called the {\em search space} which, for simplicity, will be assumed to be a hypercube: $S = [a,b]\otimes[a,b]\otimes\ldots \otimes [a,b]$.
In PSO, $f(x)$ is sampled at a fixed number of points and the coordinates of these points 
are evolved iteratively. Each point is called a {\em particle} and the set of particles is called a {\em swarm}.  At each iteration step, indexed by an integer $k = 0,1,\ldots$, the fitness is
evaluated at the current location of each particle.

Let $x_i (k)$, $i = 1, 2,\ldots,N_{\rm part}$, be the
position of the $i^{\rm th}$ particle in a swarm of $N_{\rm part}$ particles at the iteration step $k$. The coordinates corresponding to $x_i(k)$ are $(x_{i,1}(k),\ldots,x_{i,n}(k))$. 
The evolution equations for the swarm mimic in a simple way the 
behavior of real biological swarms (e.g., a flock of birds).
Associated with each particle is a memory of the location where it found the best fitness value over its past history. This location, $p_i(k)$, is called {\em pbest} (``particle best").
\begin{eqnarray}
f\left(p_i(k)\right) & = & \min_{j = k,k-1,\ldots,0} f\left(x_i(j)\right)\;.
\label{updatepbest}
\end{eqnarray}
Associated with the swarm is a memory of the location where the swarm found the
best fitness value over its past history. 
This location,  $g(k)$, is called {\em gbest} ("global best").
\begin{eqnarray}
f\left(g(k)\right) & = & \min_{j = 1,\ldots,N_{\rm part}} f\left(p_j(k)\right)\;.
\label{updategbest}
\end{eqnarray}
Given $x_i(k)$, $p_i(k)$ and $g(k)$, the following equations are used
to evolve the swarm.
\begin{eqnarray}
x_i(k+1) & = & x_i(k) +v_i(k);\label{positionupdate}\\
v_{i,j}(k+1) & = & \min\left(\max \left(y_{i,j}(k+1),-v_{\rm max}\right),v_{\rm max}\right)\;,\\
y_i(k+1) & = & w(k) v_i(k) + {\bf m}_{i,1}(p_i(k)-x_i(k))+
{\bf m}_{i,2}(g(k) - x_i(k))\;,\label{actualpsovelocity}
\end{eqnarray}
where $v_i(k)$ is called the ``velocity" of a particle.  The second and
third terms in Eq.~\ref{actualpsovelocity}, called ``acceleration" terms, change
the velocity in a random manner: ${\bf m}_{i,p}$, $p = 1,2$, is a diagonal matrix, ${\rm diag}(m_{p,i,1},\ldots,
m_{p,i,n})$, such that $m_{p,i,k}\sim U[0,c_p]$ is drawn from a
uniform distribution over $[0,c_p]$, where $c_p$ is a user
specified parameter.
The parameter $w(k)$ is called the ``inertia" of
a particle and it can change as the iteration progresses according to some specified 
law. At the termination of PSO, the highest fitness value found by the swarm, and the location of the particle with that fitness, make up the solution to the optimization problem. 

The swarm is initialized by $x_{i,j}(0)=U[a,b]$ and $v_{i,j}(0) = U[-v_{\rm max}^\prime, v_{\rm max}^\prime]$. Usually, $v_{\rm max} = v_{\rm max}^\prime$, but we fix them independently in this paper in order to promote greater exploration of
$S$ (see below). The use of the limiting speed $v_{\rm max}$ prevents the swarm from exploding and leaving $S$. The same can be accomplished by a velocity constriction factor~\citep{clerc2002particle}. 

The physical meaning of the dynamical equations is fairly easy to grasp. In the absence of the acceleration terms, each particle simply moves in a straight line set by the vector $v_i(k=0)$. With the acceleration terms on, the particle is deflected on the average  towards {\em pbest} and {\em gbest}. Thus, each particle explores the search space 
under the competing pulls of moving independently of the swarm, which encourages exploration of the search space, and moving 
towards the best location found by the swarm, which encourages convergence (or {\em exploitation}). 

In general, the behavior of the swarm transitions from exploration in the early 
phase to  exploitation and convergence to an optimum in the late phase. A longer time spent exploring leads to a better chance of locating the global optimum but it also increases the computational cost of the method. Too soon a transition to exploitation may lead to premature convergence to a local optimum. The relative time spent in the two phases should be governed by
 the nature of the fitness function. However, since the degree of multi-modality is often 
unknown, it is best to err on the side of caution and extend the exploration phase.

One way to extend the exploration phase is to
 identify a set of neighbors for each particle and use the best
value found within this neighborhood, called {\em lbest} (``local best"), in Eq.~\ref{actualpsovelocity} instead of {\em gbest}. By making each particle a member of 
multiple neighborhoods, information about {\em gbest}  leaks
 across the swarm but the rate at 
which this happens can be slowed down significantly compared to the case where each
particle is constantly aware of {\em gbest}. There are many schemes for selecting neighborhoods 
in the PSO literature with the simplest being the {\em ring topology}: the particle indices are put on a circle and the neighborhood of each particle consists of $(m-1)/2$ particles
on each side with $m$ being the user specified size of each neighborhood.

The settings for most of the parameters of the PSO algorithm outlined above
have been found to be quite robust 
across a wide range of fitness functions~\citep{bratton2007defining}. This is, in fact, one of the attractive features of PSO 
as it considerably lessens the effort needed for tuning the algorithm.
In the present paper we choose the following settings: $N_{\rm part} = 40$, $c_1 = c_2=2.0$,  $m = 3$, $v_{\rm max} = (b-a)/5$, $v_{\rm max}^\prime = (b-a)/2$, $w(k) = 0.9 - 0.5(k/(N_{\rm iter}-1))$, where $N_{\rm iter}$ is the total number of iterations. In addition to fixing the PSO parameters, the behavior of particles crossing the boundary of $S$ must be prescribed. We use the so-called ``let them fly" boundary condition in which the fitness of the particle is simply set to $-\infty$ while it is outside $S$. This ensures that boundary crossing particles continue to behave according to the PSO dynamical equations while eventually bring drawn back into $S$ after a small number of iterations..

Although successful convergence to the global
maximum is not guaranteed for any finite value for $N_{\rm iter}$, increasing $N_{\rm iter}$ increases the probability of success. 
Thus, $N_{\rm iter}$ should
be chosen to make the probability of success sufficiently close to unity while 
keeping computational costs within limits. This apparently
straightforward task is complicated, however,
 by the fact that the value of the
global maximum of $\Lambda$ is not known {\em a priori} and, hence, we do not know when PSO has succeeded. This problem can be overcome to some extent for simulated data realizations 
 following~\cite{2012AstRv...7d...4M}.  A PSO run is declared successful if it finds a 
value of $\Lambda$ that is better than the one at the known location of the true 
signal in the search space.
The underlying idea here is that  such a condition should hold for any good parameter estimation algorithm since, otherwise, estimation would never incur an error due to noise. 
This criterion can be used to 
tune $N_{\rm iter}$ using simulated data realizations. One then hopes that the same setting would also lead to a large probability of success for 
real data provided that it is well modeled by the simulations. Following this criterion, we 
found that $N_{\rm iter} = 2000$ was sufficient to give a fairly high probability of convergence for the simulations described below.
As we discuss in Sec.~\ref{sec:multiPSO}, this is a reasonable but not a 
foolproof choice since a small fraction of cases do lead to failure according to this criterion. 
However, it is possible to overcome these failures by running PSO multiple 
times on the same data realization. This computationally expensive strategy 
is not followed for the bulk of our simulations in this paper although it 
is clearly an option that should be used in any analysis of real data.

\section{Applications}\label{sec:results}

We apply the algorithm described above to simulated data from a PTA using the 
parameters of eight pulsars from the NANOGrav catalog \citep{2013ApJ...762...94D}. 
In all the cases considered below, the source is assumed to be a circular 
binary located at right ascension $\alpha=1.985~\text{rad}$ ($7~\text{hr}~35~\text{min}$) 
and declination $\delta=0.625~\text{rad}$ ($35.83^\circ$), having an orbital angular frequency 
$\omega=8.0~\text{rad~yr}^{-1}$ ($\omega_{gw}=16.0~\text{rad~yr}^{-1}$), 
and initial phase $\varphi_0=1.6~\text{rad}$ ($91.67^\circ$) at the start of 
observations. We set the inclination angle $\iota=0^\circ$, 
leading to a circularly polarized GW tensor, and the polarization angle $\psi=45^\circ$. 
The duration of the simulated observation is 5 years, with uniform 
biweekly cadence (number of samples $N_I=N_s=128$ for each pulsar). 
The noise-free timing residual (the {\em signal}) induced by this GW 
source is calculated for each pulsar in the PTA following Eq.~\ref{eq:residual2}. 
Pseudo-random sequences of white Gaussian noise with a standard deviation 
$\sigma_n=10^{-8}~\text{sec}$ are added to the signal calculated for each 
pulsar to generate realizations of PTA data. To characterize the strength 
of the signal in the data, we use the network SNR of the signal defined as 
\begin{equation}\label{eq:snr}
\rho_n=\left(\sum_{I=1}^{N_p}\langle s^I|s^I \rangle _I\right)^{1/2}
=\frac{1}{\sigma_n}\left(\sum_{I=1}^{N_p}\sum_{k=1}^{N_s}(s^I_k)^2\right)^{1/2} \,.
\end{equation}
For studying the statistical properties of the detection statistic and the 
estimated parameters, three different values of $\rho_n$ are chosen and 
500 independent realizations of data are generated for each value of $\rho_n$. 
Table~\ref{tab:table4cases} shows the values of $\rho_n$ used, and 
corresponding to each, mean and standard deviation of the astrophysically 
interesting parameters ($\alpha$, $\delta$, $\omega_{gw}$). It also lists the 
corresponding recovered SNR $\rho_r$ calculated by replacing the $s^I_k$ 
in Eq.~\ref{eq:snr} by the reconstructed signal. 
Further details about each scenario are reported separately in the following subsections.

\begin{table}[h]
\begin{center}
\begin{tabular}{c|c|cc|cc|cc|cc}
\hline\hline
Scenario & $\rho_n$ &\multicolumn{2}{|c|}{$\alpha$} & \multicolumn{2}{|c|}{$\delta$} 
& \multicolumn{2}{|c|}{$\omega_{gw}$} & \multicolumn{2}{|c}{$\rho_r$} \\
& & mean & std. & mean & std. & mean & std. & mean & std. \\

\hline
strong & 100 & 1.91 & 0.121 & 0.54 & 0.140 & 16.007 & 0.0088 & 100.1 & 1.05 \\       
            
moderate & 10 & 1.85 & 0.522 & 0.32 & 0.460 & 15.999 & 0.0844 & 10.7 & 1.00 \\

weak & 5 & 2.15 & 1.004 & 0.20 & 0.672 & 15.184 & 0.1999 & 6.4 & 0.77 \\    
\hline\hline
\end{tabular}
\end{center}
\caption{The mean and the standard deviation of the three parameters $\alpha$, $\delta$ 
and $\omega_{gw}$ calculated from 500 realizations of the three scenarios with different $\rho_n$, 
and the corresponding recovered SNR $\rho_r$.}
\label{tab:table4cases}
\end{table}

In order to detect or set up upper limits, one needs to know the distribution 
of the detection statistic under $\mathcal{H}_0$ and $\mathcal{H}_{\lambda}$. 
This involves finding the distribution of the extremum (maximum) of the 
likelihood ratio $\Lambda$, which is, in general, difficult to obtain. 
In this work we choose to use the Monte-Carlo simulation with 500 realizations 
for the $\mathcal{H}_{\lambda}$ mentioned above and 1000 realizations for 
$\mathcal{H}_0$ to obtain the respective distributions. The distribution for 
$\mathcal{H}_0$ can not be fitted by a known simple PDF, considering the distribution 
for the $\rho_n=5$ case can be fitted by a Rician distribution $R(\nu,\sigma)$, with 
the noncentrality parameter $\nu=20.1$ and the scale parameter $\sigma=5.14$. 
It turns out that the distribution converges to a normal distribution 
with the signal strength increases: for the $\rho_n=10$ case, the distribution 
can be fitted by $\mathcal{N}(\mu=57.5,\sigma=10.7)$; for the $\rho_n=100$ case, 
$\mathcal{N}(\mu=5006.3,\sigma=105.6)$.

\begin{figure}[H]
\centerline{\includegraphics[width=0.6\textwidth, angle=0]{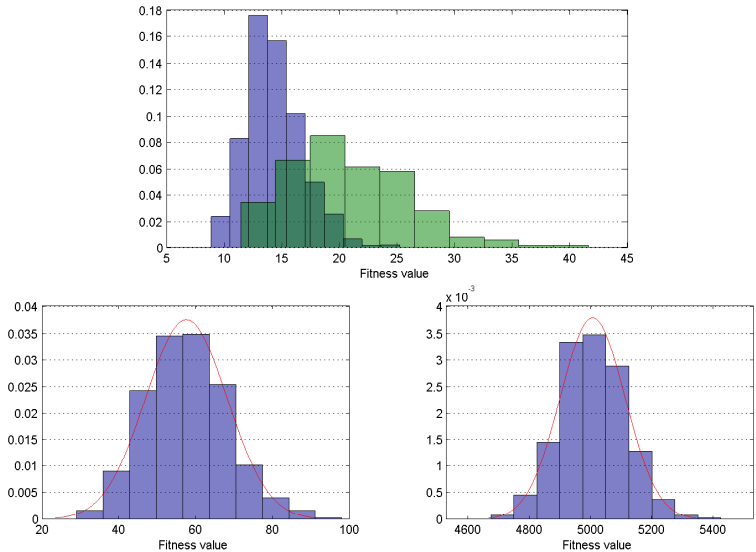}}
\caption{Histograms of the detection statistic. Blue histogram in upper panel 
is for $\mathcal{H}_0$ case, green histogram in the same panel is for $\rho_n=5$ case. Histogram 
in lower left panel is for the $\rho_n=10$ case. Histogram in lower right panel is for 
the $\rho_n=100$ case. The red curves in the lower panels show the normal distribution
with the best fit parameters mentioned in the text respectively.}
\label{fig:histh0}
\end{figure}

\subsection{Strong signal}\label{sec:strong}

\begin{figure}[H]
\centerline{\includegraphics[width=1.0\textwidth]{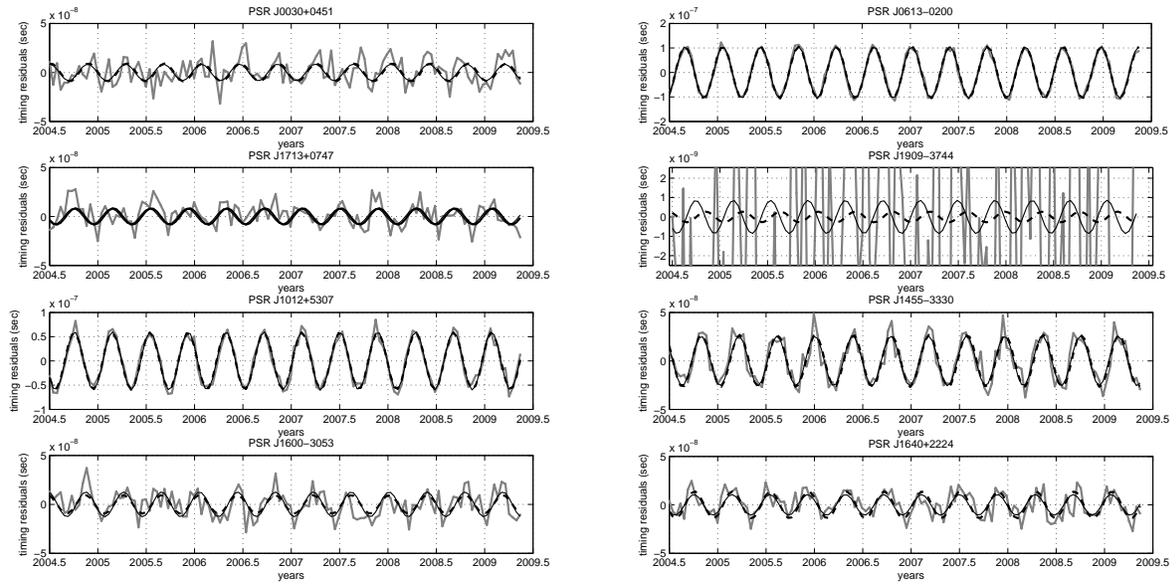}}
\caption{A data realization showing the simulated timing residuals (thick gray) 
and signal (dash black) for all pulsars. The network signal to noise ratio is $\rho_n=100$. 
The reconstructed signals are shown as solid curves. For most pulsars, the 
true and reconstructed signal are almost indistinguishable from each other.}
\label{fig:snr3_rlz127}
\end{figure}

The case $\rho_n=100$ could arise, for example, from a SMBHB with the 
chirp mass $\mathcal{M}_c\approx 10^9~M_{\odot}$, at a distance 
$D\approx 10$ Mpc. The orbital period $P=0.785$ yrs. Fig.~\ref{fig:snr3_rlz127} 
shows a typical realization of the simulated timing residuals for the eight 
pulsars (thick gray). As we can see, the magnitude and the phase of the 
noise-free timing residual (black dash) is different for the different 
pulsars in the PTA. The signal in most of the pulsars is comparable 
to or even stronger than the noise. To obtain the reconstructed signals (solid curve), 
we take the estimated intrinsic parameters and derive the extrinsic 
parameters by Eq.~\ref{eq:extrin}. Using the estimated intrinsic 
and extrinsic parameters, the estimated signal is found from 
Eq.~\ref{eq:residual2}. The estimated signal agrees with the injected 
signal very well for most pulsars, except for PSR J1909-3744, which 
has a signal amplitude about two orders of magnitude smaller than the 
others. This pulsar is almost (oppositely) aligned with the GW source 
(separation angle $\theta=174^{\circ}$), therefore it is insensitive to 
this source and has insignificant contribution to the MLR statistic. 
The mean of the recovered SNR in Table~\ref{tab:table4cases} is larger 
than the network SNR means that the detection statistic performs well in recovering the signals. 
Fig.~\ref{fig:histh0} shows the distribution of the detection statistic 
values for the signal present and absent cases. From the large separation 
of these distributions, it is clear that the detection probability $Q_d$
is nearly unity if the threshold is chosen to be the highest value found 
for the signal absent case. (The false alarm probability for this choice 
of threshold is $\simeq 10^{-3}$.)

\begin{figure}[H]
\centerline{\includegraphics[width=0.9\textwidth]{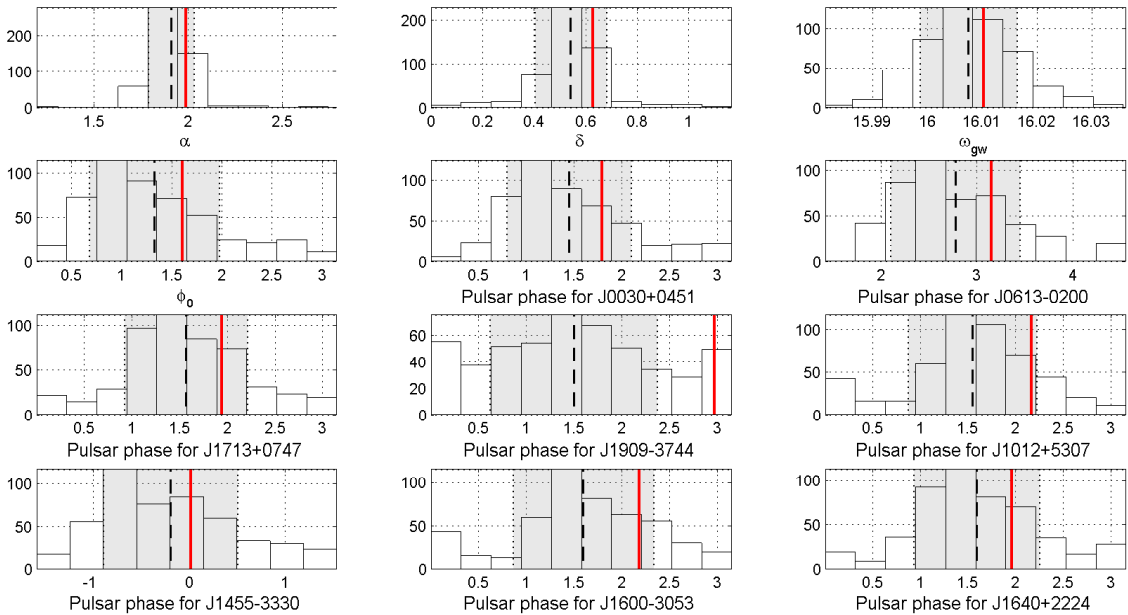}}
\caption{Histograms of the estimated intrinsic parameters for the case 
with the network signal to noise ratio $\rho_n=100$. The red vertical line 
marks the true value of the parameter used in the simulations, the dash vertical 
line marks the mean value, and shaded area covers the one sigma uncertainty. 
The total number of trials is 500.}
\label{fig:hist_paraest2}
\end{figure}

Fig.~\ref{fig:hist_paraest2} shows the distributions of the estimated intrinsic parameters 
$\{\alpha,\delta,\omega_{gw},\varphi_0,\varphi_I\}$ $(I=1,2,...,N_p=8)$ from the 500 realizations. 
The red vertical line marks the true value of the parameter used in the simulations. 
The dash vertical line marks the mean value. 
It appears that the right ascension, the declination and the frequency of GW can be 
accurately estimated with the standard deviations $\sigma_{\alpha}\approx 7^{\circ}$, 
$\sigma_{\delta}\approx8^{\circ}$, and $\sigma_{\omega_{gw}}<0.01~\text{rad~yr}^{-1}$. The one sigma contour 
of the source location distribution encloses an area of $\sim 40~\text{deg}^2$  on the sky. 
In contrast the initial orbital phase and the pulsar phases are poorly estimated even 
for this strong signal scenario. Therefore, the pulsar phase parameters are truly 
nuisance parameters that should either be maximized or marginalized over when 
constructing detection algorithms. It is not practical to estimate the pulsar 
distance from the analysis of continuous gravitational waves. 
For PSR J1909-3744, the distribution of the pulsar phase is less concentrated 
than the other pulsars due to the reason mentioned above. 
It is important to note that the initial orbital phase $\varphi_0$ and the 
pulsar phases $\varphi_I$ are directional (circular) variables, and 
their distributions should be wrapped around at the end points $0$ and $\pi$. 
For example, in the histogram of the pulsar phase 
for PSR J0613-0200, the counts for values of $\varphi_I$ below $1.5$ radians should 
be appended after the histogram ending at $\varphi_{I}=\pi$. 
Although right ascension and declination are also circular variables, 
the distributions are narrow and centered on values not close to the end points of the 
range $[0,\pi]$, therefore the wrapping of the histograms can be ignored.


\subsection{Moderate signal}\label{sec:moderate}

\begin{figure}[H]
\centerline{\includegraphics[width=1.2\textwidth]{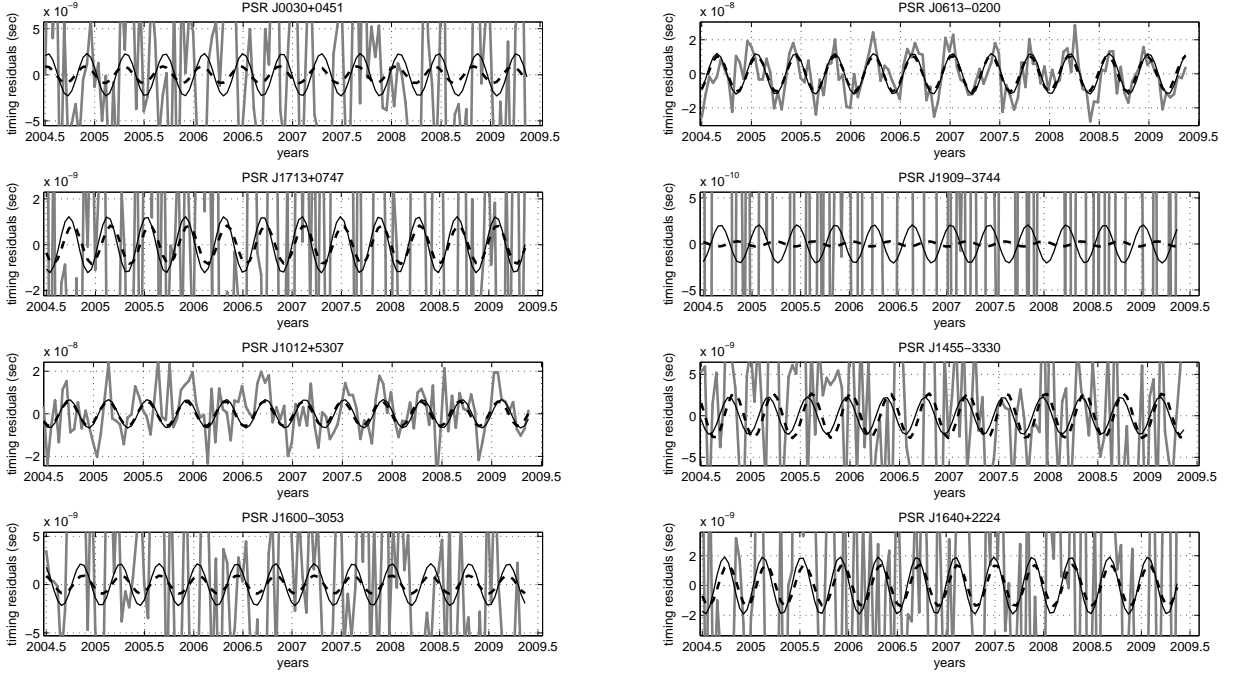}}
\caption{A data realization showing the simulated timing residuals (thick gray) 
and signal (dash black) for all pulsars. The network signal to noise ratio is $\rho_n=10$. 
The reconstructed signals are shown as solid curves. For some pulsars, such as 
PSR J0030+0451 and J1909-3744, we have zoomed into the noise in 
the subplots, so that the signal can be manifested. 
}
\label{fig:snr2_rlz247}
\end{figure}

\begin{figure}[H]
\centerline{\includegraphics[width=0.9\textwidth]{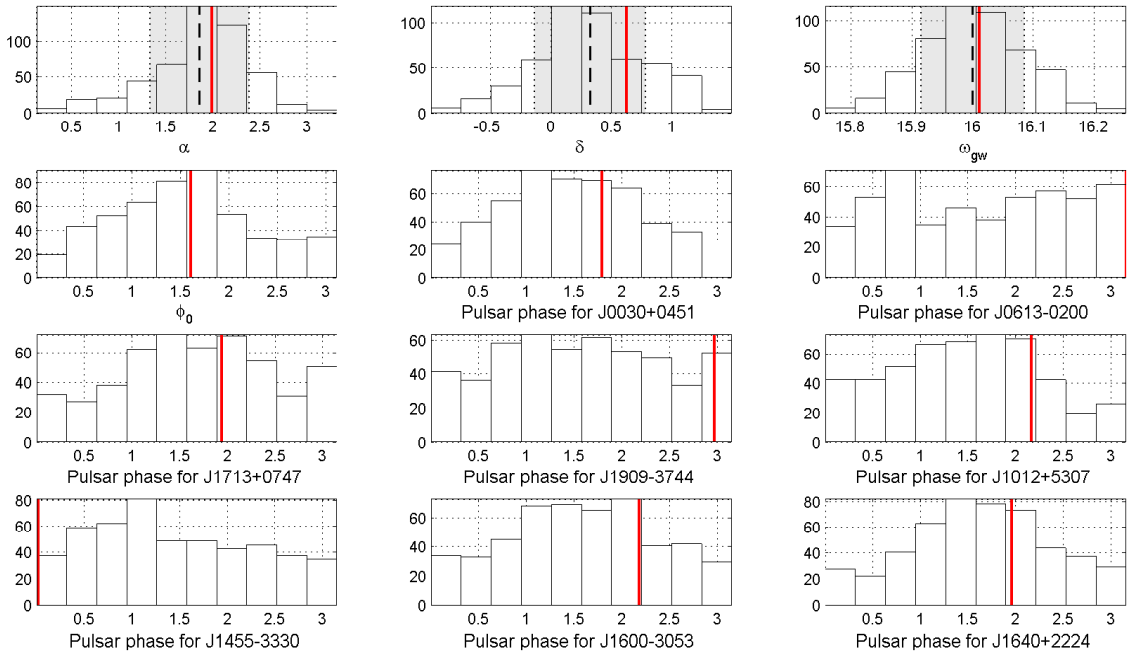}}
\caption{Histograms of the estimated intrinsic parameters for the case 
with the network signal to noise ratio $\rho_n=10$. The red vertical line 
marks the true value of the parameter used in the simulations. 
The total number of trials is 500.}
\label{fig:snr2parahist}
\end{figure}

In this scenario, the network signal to noise ratio $\rho_n=10$. This corresponds, 
for example, to a SMBHB with chirp mass $\mathcal{M}_c\approx10^9~M_{\odot}$ at 
a distance $D\approx100$ Mpc. Fig.~\ref{fig:snr2_rlz247} shows a realization 
of the simulated timing residuals for this scenario. The recovered signal 
agrees well with the injected signal in 5 of the pulsars that have the 
strongest signals, comparing the others having various levels of offsets in phase 
and changes in amplitude. Although the estimated and true signals now agree 
well in only about 5 of the pulsars, the recovered SNR (c.f., Table~\ref{tab:table4cases}) 
is still high. This indicates that a strong detection is still possible in 
this scenario, but estimation accuracy has worsened. 
Fig.~\ref{fig:snr2parahist} shows the distributions of the estimated intrinsic parameters. 
The estimation of the sky location is considerable worse than the $\rho_n = 100$ scenario, 
with the standard deviations $\sigma_{\alpha}\approx 30^{\circ}$, $\sigma_{\delta}\approx 26^{\circ}$ 
respectively. The one sigma contour of the source location distribution encloses an 
area of $\sim 800~\text{deg}^2$  on the sky. However the GW frequency can still be 
estimated quite accurately with $\sigma_{\omega_{gw}}<0.1~\text{rad~yr}^{-1}$.

\subsection{Weak signal}\label{sec:weak}

\begin{figure}[H]
\centerline{\includegraphics[width=1.2\textwidth]{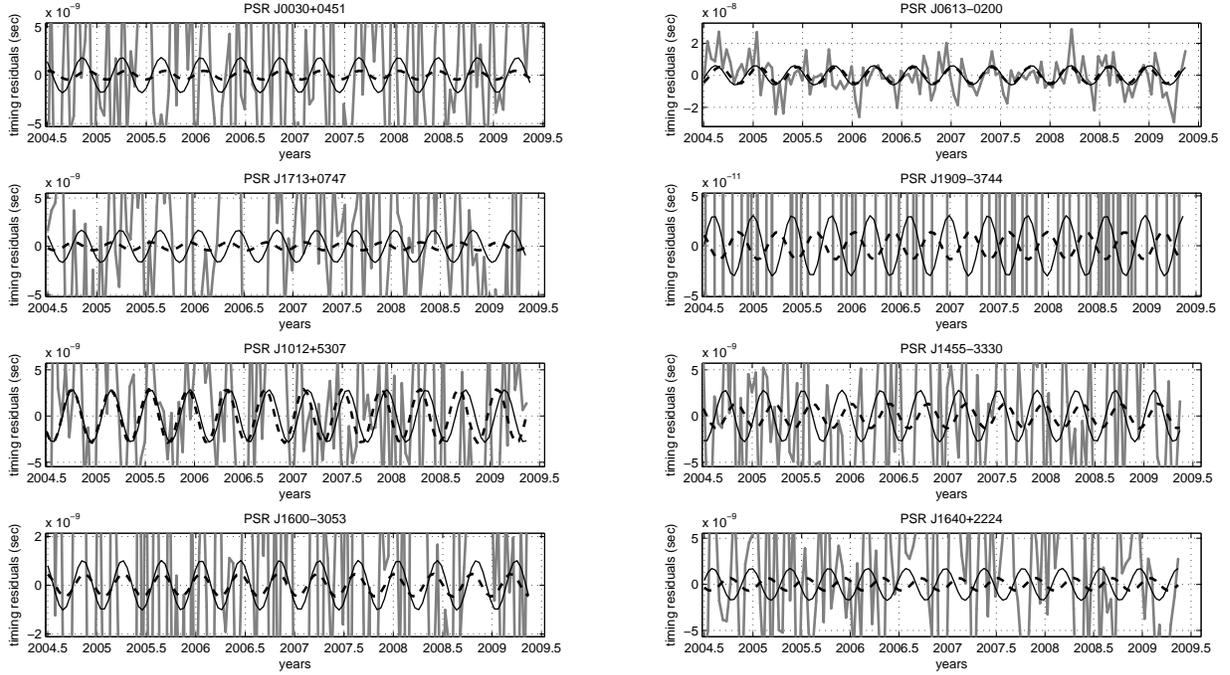}}
\caption{A data realization showing the simulated timing residuals (thick gray) 
and signal (dash black) for all pulsars. The network signal to noise ratio is $\rho_n=5$. 
The reconstructed signals are shown as solid curves. For all pulsars, we have zoomed 
into the noise in the subplots, so that the signal can be manifested.}
\label{fig:snr1_rlz122}
\end{figure}

\begin{figure}[H]
\centerline{\includegraphics[width=0.9\textwidth]{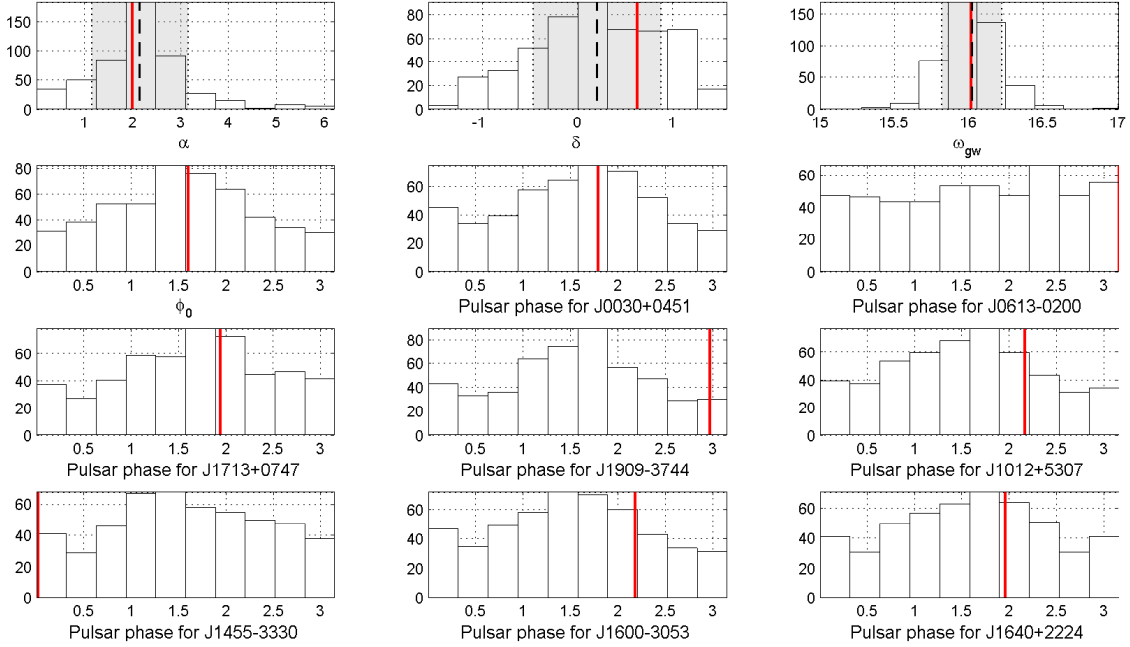}}
\caption{Histograms of the estimated intrinsic parameters for the case 
with the network signal to noise ratio $\rho_n=5$. The red vertical line 
marks the true value of the parameter used in the simulations. 
The total number of trials is 500. }
\label{fig:snr1parahist}
\end{figure}

In this scenario, the signal strength is weak with a network signal to noise 
ratio $\rho_n=5$. This corresponds, for example, to a SMBHB with chirp 
mass $\mathcal{M}_c\approx10^9~M_{\odot}$ and the distance $D\approx10^3$ Mpc.
Fig.~\ref{fig:snr1_rlz122} shows a realization of the simulated timing residuals 
in which the noise is seen to dominate the signal for all the pulsars. 
This realization illustrates the most likely case for the current level of 
sensitivity for gravitational wave detection using pulsar timing arrays. 
Fig.~\ref{fig:snr1parahist} shows the distribution of the estimated intrinsic 
parameters. Consequently, the reconstructed signals for only two pulsars 
agree reasonably well with the true signals, whereas for most of the pulsars the match is quite poor. 
From Fig.~\ref{fig:histh0}, the detection probability $Q_d\simeq 18.6\%$ if we choose 
the threshold to be the highest value obtained for the noise-only case. 
The one sigma contour encloses an area more than $10^3~\text{deg}^2$, a large fraction of 
the sky. The estimation of GW frequency is tolerable. Therefore in this case, although the 
signal is still strong enough to be detected, it is not strong enough to be localized.

\subsection{Improvements from multiple runs of PSO}\label{sec:multiPSO}

\begin{figure}[H]
\centerline{\includegraphics[width=1.0\textwidth, angle=0]{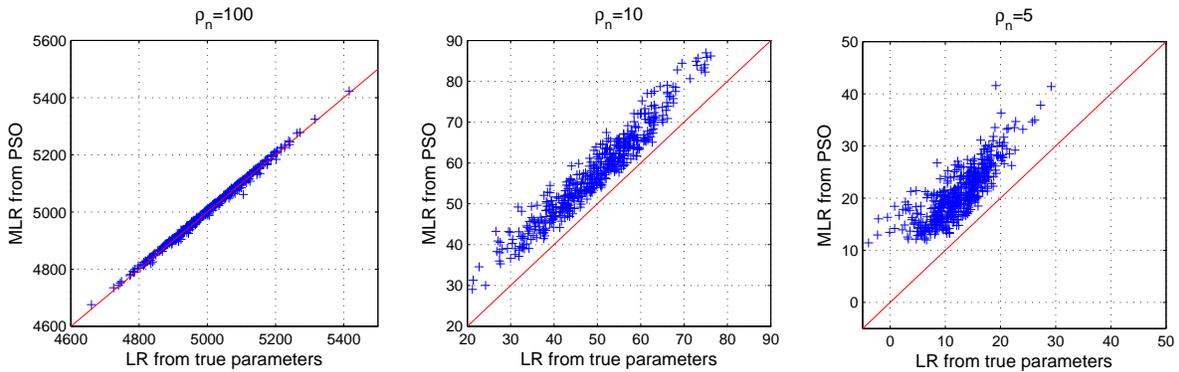}}
\caption{MLR calculated with estimated parameters versus LR calculated with true parameters 
for the 500 realizations for the $\rho_n=100$, 10 and 5 scenarios respectively.}
\label{fig:MLRvsLR}
\end{figure}

As mentioned in Sec.~\ref{sec:pso}, we tuned the number of iterations ($N_{iter}$) 
in PSO such that it converges to the global maximum for the bulk of the trials 
in our simulations. The evidence for a successful convergence is the achievement 
of a better fitness value than the one at the known location of the true signal. 
Fig.~\ref {fig:MLRvsLR} shows the fitness value calculated with the estimated 
parameters (MLR) using the algorithm and PSO versus the fitness value calculated 
with the true parameters (LR) used for the 500 realizations of the simulated data. 
The blue cross above the red diagonal line means that the MLR is larger than the LR 
for the realization. As we can see, this is true for all realizations for 
$\rho_n=10$ and $\rho_n=5$ cases, comparing about $20\%$ of the realizations 
for $\rho_n=100$ are below the red line.

It is possible to remedy the small fraction of failures seen in Fig.~\ref{fig:MLRvsLR}
by a simple strategy: perform multiple independent runs of PSO on the same data 
realization and use the output from the run that finds the best fitness value. 
Although it makes perfect sense to use this strategy in any implementation involving 
real data, it is computationally too expensive for a simulation involving hundreds 
of data realizations. Here we only demonstrate the viability of this strategy 
by applying it to a small fraction of the simulated data realizations.

\begin{figure}[H]
\centerline{\includegraphics[width=1.0\textwidth, angle=0]{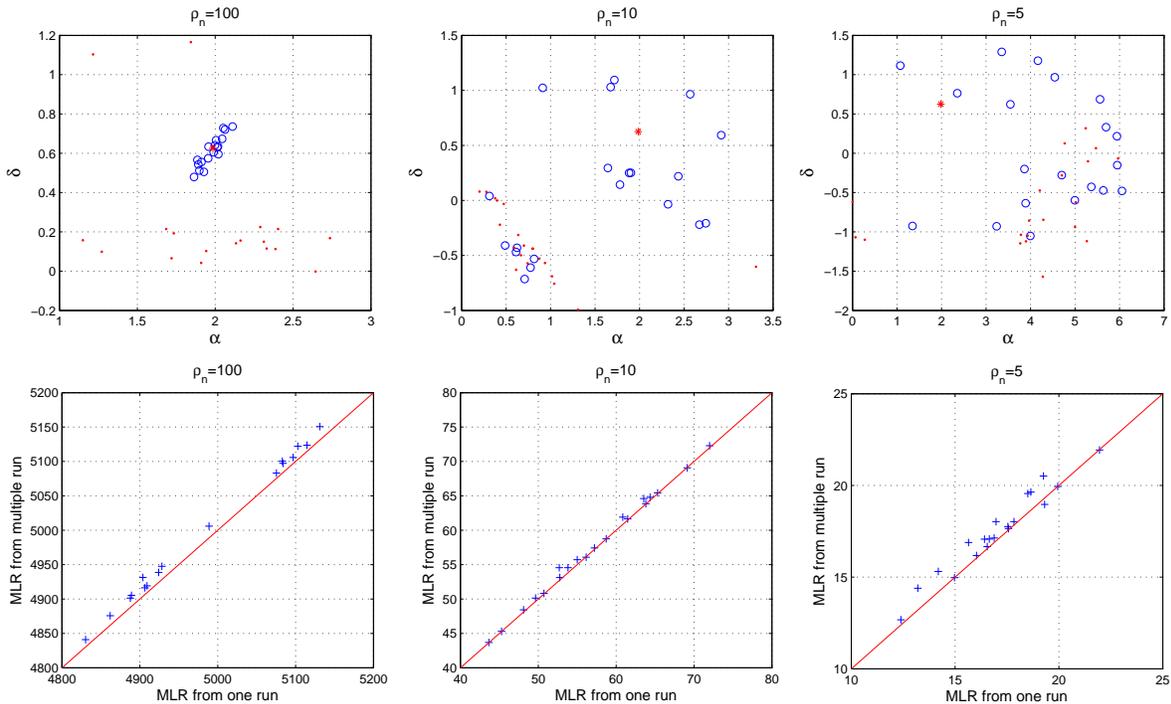}}
\caption{In each column, the top subpanel shows estimated sky locations, 
with results from the rerun of the method (10 independent PSO runs) shown 
as circles and the results from the earlier runs (1 PSO run) shown as dots. 
The X-axis shows right ascension and Y-axis shows declination. The true 
location is shown as an asterisk. The bottom panel in each column shows 
the MLR from this method versus the MLR from the earlier runs.}
\label{fig:ReRunXtreme}
\end{figure}

For each scenario in Table~\ref{tab:table4cases}, we select $4\%$ of the realizations 
for which the estimated sky locations departed most from the true location. 
We then rerun the detection and estimation method, with 10 independent runs of 
PSO instead of one, on each of these data realizations, and use the results from 
the run that terminates with the best fitness value.  Fig.~\ref{fig:ReRunXtreme} 
compares the estimated sky locations and fitness values found by the old and new runs. 
For the $\rho_n=100$ case, the estimated sky locations clearly improve significantly. 
For the $\rho_n=10$ and $\rho_n=5$ cases, the improvement is not as dramatic. 
However, we see that the new estimates are dispersed around the true location 
in a much more symmetrical way than the previous ones. Although this does not 
improve the variance of the estimates, it clearly reduces bias.
For all values of $\rho_n$, the MLR is improved for most data realizations. 
Note that the new set of runs of PSO are independent of the old ones as they start 
off with different initial conditions. Hence, it is not surprising that the new 
strategy may perform marginally worse in some rare cases. Again, increasing 
the number of independent PSO runs a bit more should considerably diminish the 
chances of this happening. We also see that the typical increase of MLR is 
smaller than the bin size of the histograms in Fig.~\ref{fig:histh0}. 
Therefore the distribution of the detection statistic will 
not change considerably if we use multiple runs of PSO for all the realizations. 
Estimation performance of the method, however, should improve somewhat. 
(The significant improvement for the $\rho_n=100$ case is tempered by the 
fact that it represents an extremely unlikely signal strength.)

\section{Conclusions} \label{sec:sum}

We have presented a coherent network analysis method based on the GLRT for 
the detection and estimation of monochromatic continuous gravitational 
waves using a pulsar timing array. The method explicitly includes pulsar 
phase parameters in the maximization of the likelihood ratio. The resulting 
high dimensional optimization problem is successfully addressed using 
Particle Swarm Optimization. By keeping the pulsar phases as intrinsic 
parameters in the maximization, the method resembles the $\mathcal{F}$-statistic 
in the choice of extrinsic parameters. However, our results show that 
the pulsar phases are uninformative and should be treated as nuisance 
parameters in future studies. Maximization or marginalization over these 
parameters will result in a detection statistic that is radically different 
from the $\mathcal{F}$-statistic. Quantifying the performance of this 
statistic will be the subject of future work. 

Even if one treats the pulsar phases as extrinsic parameters, the number of 
intrinsic parameters will not be small. Hence, stochastic optimization methods 
such as PSO will still be required. We have shown that PSO already works well 
with only minimal effort required in its tuning. Therefore, we are confident 
that it will continue to be useful when we shift to a method that treats pulsar 
phases as extrinsic parameters.

A limitation of the present study is that the simulated observations are 
evenly sampled with biweekly cadence, and have stationary white Gaussian 
noise with zero mean and the same variance for all pulsars. However, since 
the method presented here works entirely in the time domain, no modifications 
are needed to switch to irregularly sampled data. Given that the correct 
way to proceed appears to involve changing the  pulsar phases into extrinsic 
parameters, simulations with more realistic data models are best postponed 
until such a method is realized. \citet{2001PhRvD..63j2001F} has shown already 
that coherent methods, such as the one presented here, are robust against 
non-Gaussianity in timing residual noise. Thus, the Gaussian noise assumption 
in our simulations, although incommensurate with expectations about real 
data \citep{2014inPrepYWANG}, is not a serious flaw.

So far, we have assumed that the timing residuals are a superposition of the signal 
and the noise. In reality, the timing residuals are obtained by a weighted least 
square fit as mentioned in Sec.~\ref{sec:intro}. Due to the covariance of the GW signal  
with some of the fitting parameters, the power of GW signal may be absorbed in the fitting 
model, and the signal in the residuals may be changed. Besides, the fitting 
procedure can also change the statistics (e.g. covariance matrix) of the noise, especially the 
noise component can become non-stationary in the timing residual even though the 
noise in the TOA is stationary. 
In order to detect gravitational waves, one must take account of these fitting 
effects. One approach is to use the projection operator (matrix) $\mathbf{R}$ 
define in \citet{2013ApJ...762...94D}. It only depends on the fitting model and 
the weighting matrix, not the specific value of the data. This allows us 
to study the fitting effects by simulations. 
These factors should be taken into account in the analysis when we apply 
this algorithm to the real data. 

Another element required in real data analysis but absent from our 
study is noise characterization. In this paper, we assumed that the noise parameters 
are known \textit{a priori} or can be fixed at their independently 
estimated values. A more comprehensive approach would be to include 
them as part of the estimation process along with the signal 
parameters. Though straightforward in terms of the formalism, 
this will increase the dimensionality of the search space for PSO significantly 
(e.g., from 12 to 52 for a PTA of 8 pulsars). PSO is routinely used for 
optimization problems with comparable dimensionality, but it remains to be 
seen if the standard variant of PSO used in this paper will continue to work 
successfully when the above extension is made.

The method we have presented does not account for elliptical orbits or 
evolution of the orbital parameters. It is expected that various 
dissipation effects (including GWs) would have circularized the orbit 
by the time it enters the sensitive frequency band of PTA. 
The evolution of orbital parameters becomes a progressively important 
consideration as the signal frequency, linked to the orbital frequency, 
increases. The effect of orbital evolution during the travel time of radio 
pulses to Earth is that the pulsar term corresponds to a lower orbital 
frequency (earlier stage of binary source) than the one in the Earth term. 
In addition, according to Eq.~\ref{eq:zeta}, the amplitude of the pulsar term 
will be larger than the Earth term. 
These considerations can be accomodated in the analysis by the introduction of one extra 
intrinsic parameters, which will not affect the performance of PSO. 
The estimation errors of all the parameters, however, will worsen as a result.
This work is subject to our future investigations.

\section{Acknowledgments}

This work was supported by the National Science Foundation under PIRE grand 0968296. 
The contribution of SDM to this paper is supported by NSF awards PHY-1205585 and
HRD-0734800. We are grateful to the members in the NANOGrav for helpful comments and discussions.

\bibliographystyle{apj}
\bibliography{pta4gw}
\end{document}